\documentclass[prb,twocolumn,superscriptaddress,showpacs,preprintnumbers,amsmath,amssymb]{revtex4}
\usepackage{amssymb}

\usepackage{graphicx}
\usepackage{dcolumn}
\usepackage{bm}
\usepackage{epsfig}
\usepackage{amsmath}
\newcommand{\ignore}[1]{}

\begin{document}


\title{Quantum Size Effect on Dielectric Properties of Ultrathin Metallic Film: A First-Principles Study of Al(111)}

\author{Wenmei Ming}
\affiliation{Department of Materials Science and Engineering,
University of Utah, Salt Lake City, UT 84112}

\author{Steve Blair}
\affiliation{Department of Electrical and Computer Engineering,
University of Utah, Salt Lake City, UT 84112}

\author{Feng Liu}
\thanks{Corresponding author. E-mail: fliu@eng.utah.edu}
\affiliation{Department of Materials Science and Engineering,
University of Utah, Salt Lake City, UT 84112}

\date{\today}


\begin{abstract}
Quantum manifestations of various properties of metallic thin films
by quantum size effect (QSE) have been studied intensively. Here,
using first-principles calculations, we show quantum manifestation
in dielectric properties of Al(111) ultrathin films. The QSE on the
dielectric function is revealed, which arises from size dependent
contributions from both intraband and interband electronic
transitions. More importantly, the in-plane interband transitions in
the films thinner than 15 monolayers are found to be smaller than
the bulk counterpart in the energy range from 1.5~eV to 2.5~eV. This
indicates less energy loss with plasmonic material of Al in the form
of ultrathin film. Our findings may shed light on searching for
low-loss plasmonic materials via quantum size effect.
\end{abstract}.

\maketitle

\section{Introduction}
When the dimension of a material (either one, two or all three
dimensions) is reduced to nanoscale, its properties become size
dependent, due to quantum confinement effect or surface effect, and
the former is generally referred to as the quantum size effect
(QSE)\cite{chargeden1, chargeden2}. In recent years, there have been
intensive studies of QSE on properties of ultrathin films in the
quantum confinement regime. In particular, when the thickness of a
metal film is comparable with the electron Fermi wavelength, quantum
confinement becomes prominent resulting in discretization of
electronic states (quantum-well states). This in turn leads to a
variety of strongly thickness-dependent film properties, manifested
as QSE, such as surface energy \cite{stability1, stability2},
surface stress \cite{Hu,Miao},
 surface diffusion barrier
\cite{diff1,diff2}, surface reactivity \cite{chem}, work function \cite{workf}, elastic constant \cite{Miao2} and so on.

Furthermore, considering the emerging field of plasmonics
\cite{plasma1, plasma2, plasma3} for for myriad applications, the
QSE is expected to also affect the dielectric and optical properties
of ultrathin metal films. Theoretically, Dryzek \emph{et al}
\cite{QSEopt1} and Apell \emph{et al} \cite{QSEopt2} showed QSE on
the optical spectra of gold and potassium thin films, respectively,
within finite free electron model, which differs from the classical
Drude model. Laref \emph{el al} \cite{QSEopt3} studied the thickness
dependence of optical permittivity of ultrathin gold films for a
more accurate design of plasmonic device, based on density
functional theory (DFT) calculations. Experimentally, Kuzik \emph{et
al} \cite{QSEopt4, QSEopt5} observed the periodic oscillation of
optical conductivity as a function of film thickness in Nb, Cu, Mo,
W, Ni and Ti, which is possibly an indication of QSE.

Generally, the simple metals (e.g. Ag, Au, Cu, Al and Mg) are good
candidates to generate plasma, because they have a high free carrier
density around Fermi level that leads to a large bulk plasma
frequency of about 10~eV. This makes the workable frequency of the
externally applied electromagnetic field cover a wide range from
visible to ultraviolet (UV). However, one problem with using these
metals is the large energy loss of the applied electromagnetic
field, largely because the absorption induced by interband
electronic transitions falls also into the visible and UV range
\cite{plasma1, plasma4}. Band structure engineering, therefore, is
required to remove the interband transitions or push them out of
visible and UV range, especially for UV plasmonic applications. One
interesting idea is to reduce the thickness of metal films so that
one may take the advantage of QSE to reduce the absorption in the
frequency range of interest. This idea has motivated the present
study of QSE on the dielectric properties of ultrathin Al(111)
films.

We have investigated the QSE on the dielectric function of Al(111)
films with both intraband and interband contributions in the
thickness range from 1 monolayer (ML) to 39~MLs, using
first-principles DFT calculations within the random phase
approximation (RPA) \cite{slaboptic}. We found that the plasma
frequency is reduced due to the transfer of electronic transition
strength from the intraband to interband transitions and the amount
of such reduction oscillates as a function of film thickness. Also,
we found that for the films thinner than 15~MLs, the imaginary part
of in-plane dielectric constant with energy greater than 1.5~eV is
decreased to be smaller than the bulk value. This may translate into
less energy loss in the plasmonic devices made of Al(111) films
thinner than 15~MLs.

\section{Calculation details and theory} Our DFT calculations were
performed by using projector augment wave pseudopotential (PAW)
\cite{PAW} with the generalized gradient approximation (GGA)
\cite{GGA} to the exchange-correlation functional, as implemented in
VASP package \cite{vasp}. The Al(111) slab plus a more than 20 {\AA}
vacuum was used as our model film structure. 350~eV energy cutoff
and $61\times61\times1$ $\Gamma$-centered k-mesh were used for
wavefuntion expansion and k-space integration, respectively, to
achieve a highly converged dielectric function calculation. All the
structures were relaxed in terms of internal atomic coordinates
using conjugate gradient method until the force exerted on each atom
is smaller than 0.01~eV$/{\AA}^3$. For a better bulk reference, we
used a hexagonal unit cell with three Al(111) layers without vacuum,
and the same energy cutoff and in-plane k-point sampling as the slab
calculation to calculate the bulk Al dielectric function. The
in-plane lattice constant of 2.8567~{\AA} for Al(111) film was used.

In general, the optical absorption results from three mechanisms:
plasmon resonance, free carrier intra-band damping and bound
electron inter-band transitions. The first one is responsible for
exciting plasma in plasmonic devices, but the latter two are
detrimental to the performance of plasmonic devices. In this work,
we focus on possible reduction of absorption due to the latter two
mechanisms by QSE, so the absorption from plasmon resonance is not
considered. For the absorption from plasmon resonance in
nanostructures, the nonlocal part of the dielectric function may
have a noticeable effect on the adsorption spectra because of the
plasmon-resonance induced large non-uniform electric field
distribution that cannot be described by local electrodynamics.
However, for free carrier intra-band damping and bound electron
inter-band transitions, connecting the optical adsorption to the
local description of dielectric function is generally valid as long
as the optical wavelength is much larger than the unit-cell
dimension and the thickness of the film. This approximation has been
successfully used in calculating the optical properties of low
dimensional structures, such as nanotubes \cite{nanotubes1},
graphene systems \cite{graphene1} and nanowires \cite {nanowire2}.
Therefore, we also used the local dielectric function to investigate
the optical absorption from electron intra-band damping and
inter-band transitions.

In a partially filled metal like Al, the intraband transitions
around Fermi energy need also to be included, in addition to the
interband transitions. The interband-transition contribution to the
dielectric function is in the form of 2$^{nd}$ rank tensor and its
imaginary part is expressed as \cite{opfor}:
\begin{align}\label{inter2}
\varepsilon_{\alpha\beta}^{2,inter}(\omega)=&\frac{8\pi^2e^2}{\Omega
m^2\omega^2}\underset{n\neq m;k}{\sum}(f_{mk}-f_{nk})P_{e_{\alpha};
m,n,k}P_{e_{\beta}; m,n,k} \nonumber  \\ &\times\delta
(\epsilon_{mk}-\epsilon_{nk}-\hbar \omega)
\end{align}
Where $\Omega$ is the volume of the unit cell, \emph{f$_{nk}$} is
Fermi-Dirac occupation function. e$_{\alpha}$ is the unit Cartesian
directional vector of electric field polarization and
$\epsilon_{nk}$ is the electron eigen-energy. $P_{e_{\alpha};
m,n,k}$ is the momentum matrix element between the Bloch
wavefunctions of (m, k) and (n, k) with the projection onto the
e$_{\alpha}$ direction. The corresponding real part can be computed
by Kramers-Kronig relation.

The intraband-transition contribution to the dielectric function is
usually expressed in Drude form:
\begin{eqnarray}
\varepsilon_{\alpha\beta}^{2,intra}=\pi
\omega^2_{pl;\alpha\beta}\frac{\partial\delta(\omega)}{\partial\omega}
\end{eqnarray}
$\omega^2_{pl;\alpha\beta}$ is a tensor which coincides with the
bulk plasma frequency tensor in bulk calculation and follows the
expression:
\begin{eqnarray}\label{plasma}
\omega^2_{pl;\alpha\beta}(\omega)=\frac{8\pi^2e^2}{\Omega\hbar^2m^2}\sum_{n,k}2\delta(\epsilon_{nk}-\epsilon_F)(e_{\alpha}\frac{\partial
\epsilon_{nk}}{\partial k})(e_{\beta}\frac{\partial
\epsilon_{nk}}{\partial k})
\end{eqnarray}
This equation gives the imaginary part of intraband dielectric
function within the independent particle picture, where the electron
lifetime is infinite. However, the scattering from
electron-electron, electron-phonon and electron-defect interactions
may result in a finite life-time. Thus, a lifetime broadening
parameter $\Gamma$ is often introduced in order to compare the
experimental dielectric function with the calculated one. The
equation is then modified to include this free carrier intra-band
damping:
\begin{eqnarray}\label{intra2}
\varepsilon_{\alpha\beta}^{2,intra}(\omega)=\frac{\Gamma
\omega^2_{pl;\alpha\beta}}{\omega(\omega^2+\Gamma^2)}.
\end{eqnarray}
The film volume excluding the vacuum region from the supercell was
used to normalize the dielectric function in the above equations.
The interband transitions and intraband damping induced optical
absorption energy loss per unit volume is proportional to the
imaginary part of the dielectric function \cite{optA1, optA2}.

The interband and intraband dielectric functions are coupled with
each other through the \emph{f}-sum rule as \cite{fsumrule}:
\begin{eqnarray}\label{fsum}
\int_0^{\infty}\omega[\varepsilon_{\alpha\alpha}^{2,intra}(\omega)+\varepsilon_{\alpha\alpha}^{2,inter}(\omega)]d\omega=\frac{2\pi^2e^2n\hbar^2}{m}
\end{eqnarray}
Where n is the valence electron number density. In the following
section this sum-rule will be used to explain the difference between
slab plasma frequency and bulk plasma frequency.
\begin{figure}
\includegraphics[clip,scale=0.60]{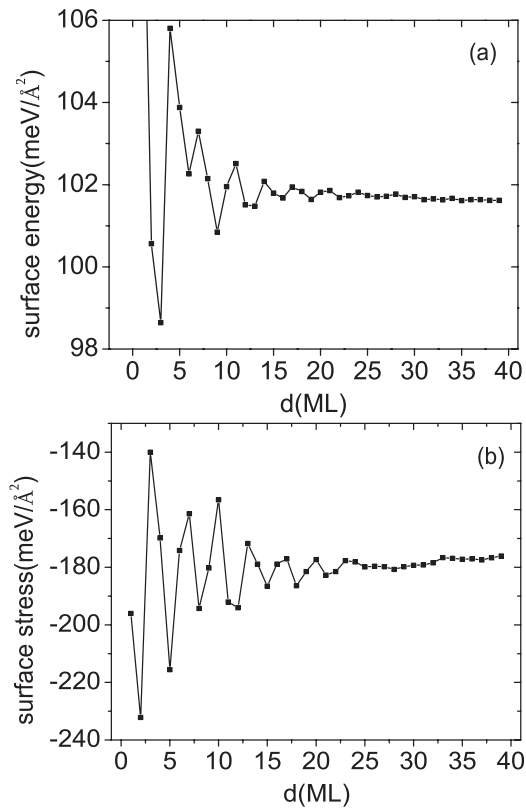}
\caption{\label{Fig1} (a) Surface energy of free-standing Al(111)
film as a function of thickness in unit of monolayer; (b) The
corresponding surface stress.}
\end{figure}

\section{Results and discussion}
Bulk Al has a high plasma frequency of about 12.4~eV and its
dielectric properties as a function of incident photo energy were
studied by previous studies \cite{Alplas1}. The most striking
feature from the imaginary part of dielectric function is that it
has two interband absorption peaks around 0.5~eV and 1.5~eV due to
the transitions between parallel bands with energy difference of
0.5~eV and 1.5~eV, respectively. For the thin film form of Al(111),
because the interlayer distance and the half electron Fermi
wavelength have a simple ratio of about 4:3, it suggests a strong
QSE \cite{QSEAl}, similar to the case of Pb(111) thin film with a
ratio close to 3:2 \cite{stability1}. We expect the quantum
manifestation in the plasma frequency and interband transitions will
appear in Al(111) ultrathin films. To illustrate the possible QSE,
we first calculated surface energy and surface stress as a function
of film thickness, as shown in Fig. 1. Clearly, surface energy in
Fig. 1(a) displays an oscillation with a 3~ML periodicity
superimposed by a 10~ML beating pattern, especially for the first
30~MLs, in good agreement with previous result \cite{FEG}. The 3~ML
periodicity corresponds to the film thickness at which the electron
forms a standing-wave and the 10~ML beating pattern arises from the
imperfect matching between interlayer distance and half Fermi
wavelength. Surface stress displays also a QSE-induced oscillation,
as showed in Fig. 1(b).

\begin{figure}
\includegraphics[clip,scale=0.60]{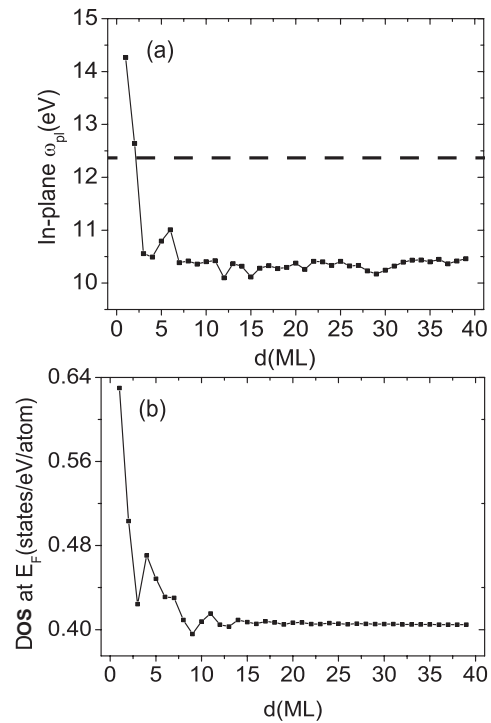}
\caption{\label{Fig2} (a) The slab plasma frequency (in eV) as a
function of Al(111) film thickness. The dashed line indicates the
bulk plasma frequency. (b) the electron density of states (DOS) at
Fermi energy as a function of film thickness.}
\end{figure}

Next, we calculated the in-plane plasma frequency as a function of
film thickness in Fig. 2(a), which shows the film thickness
dependence of intraband contribution to the dielectric function as
defined in equation (\ref{intra2}). First, there is a very fast
decay of plasma frequency from 14.3~eV to 10.5~eV going from the 1~
ML to 3~ML film. Second, there appears also a 3~ML oscillation
pattern in the plasma frequency. To understand these observations,
we noticed from equation \ref{plasma} that the squared plasma
frequency will be proportional to the electron density of states
(DOS) at the Fermi energy if $e_{\alpha}\frac{\partial
\epsilon_{nk}}{\partial k}$ varies very slowly with k or remains
constant. Therefore, we calculated the DOS at the Fermi energy at
different film thickness in Fig. 2(b). It shows a very similar
oscillation pattern as the plasma frequency in Fig. 2(a). In
particular, the DOS drops significantly from the 1~ML, 2~ML to 3~ML
film and then converges quickly to the bulk value with further
increasing film thickness. We thus attributed the large plasma
frequency of the ultrathin 1~ML, 2~ML and 3~ML films to their large
DOS at the Fermi energy. However the oscillation pattern of plasma
frequency does not match exactly in a one-to-one fashion to that of
DOS at the Fermi energy. This may be related to the modulation of
$e_{\alpha}\frac{\partial \epsilon_{nk}}{\partial k}$ term in
equation (\ref{plasma}) by QSE, which is also thickness dependent.

We also calculated the in-plane plasma frequency of bulk Al with
both 3~ML and 39~ML unit cells and obtained the same value of 12.4~
eV. However, from Fig. 2(a) the slab plasma frequency shows sizable
deviation from its bulk value even for the film thickness of 39~MLs.
The similar difference has been previously reported for Cu(110)
\cite{slaboptic} and for Au(111) \cite{QSEopt3} surfaces. It depends
on the metal species and surface orientation. Physically the
calculated in-plane plasma frequency in the slab setup should
converge to its bulk value in the limit of infinite film thickness.
This in turn means the thickness used in our calculation is still
not thick enough to reach the bulk value. Such a difference can be
qualitatively understood from the finite free-electron gas model
\cite{freegas1, freegas2, freegas3, freegas4}: with finite film
thickness, the contribution of interband transitions due to quantum
confinement induced energy level discreteness to the dielectric
function inversely depends on the film thickness, which will vanish
in the bulk limit  and leave only the commonly termed intraband
contribution to the dielectric function. With the \emph{f}-sum rule
considered then the intraband contribution will proportionally
increase towards the bulk, so that the slab plasma frequency will be
smaller than the bulk value and converge to bulk plasma frequency
with increasing thickness. On the other hand, we noticed that in
terms of dielectric function bulk Al cannot be fully described by
free electron model, because it has finite interband contribution
for all frequencies from high to even $\omega\rightarrow 0$~eV
\cite{smallw1,smallw2}, hence Al is expected to have more complex
dielectric function profile as a function of thickness than that
predicted from finite free electron gas model as we will see from
the results below.

\begin{figure}
\includegraphics[clip,scale=0.35]{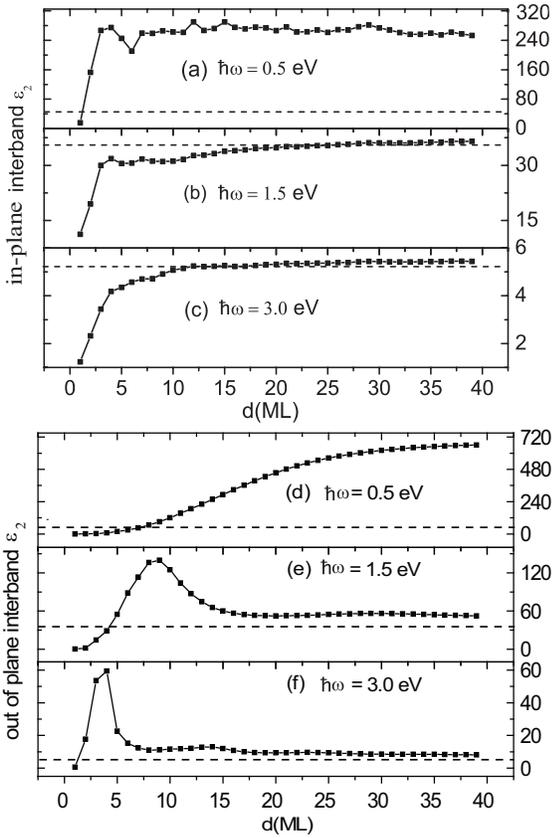}
\caption{The imaginary part of interband in-plane dielectric
function at three different energies: (a) $\hbar\omega=$ 0.5~eV; (b)
$\hbar\omega=$ 1.5~eV; (c) $\hbar\omega$= 3.0~eV. The imaginary part
of interband out-of-plane dielectric function at three different
energies: (d) $\hbar\omega=$ 0.5~eV; (e) $\hbar\omega=$ 1.5~eV; (f)
$\hbar\omega$= 3.0~eV. The dashed lines indicate the corresponding
bulk values.}
\end{figure}

\begin{figure}
\includegraphics[clip,scale=0.60]{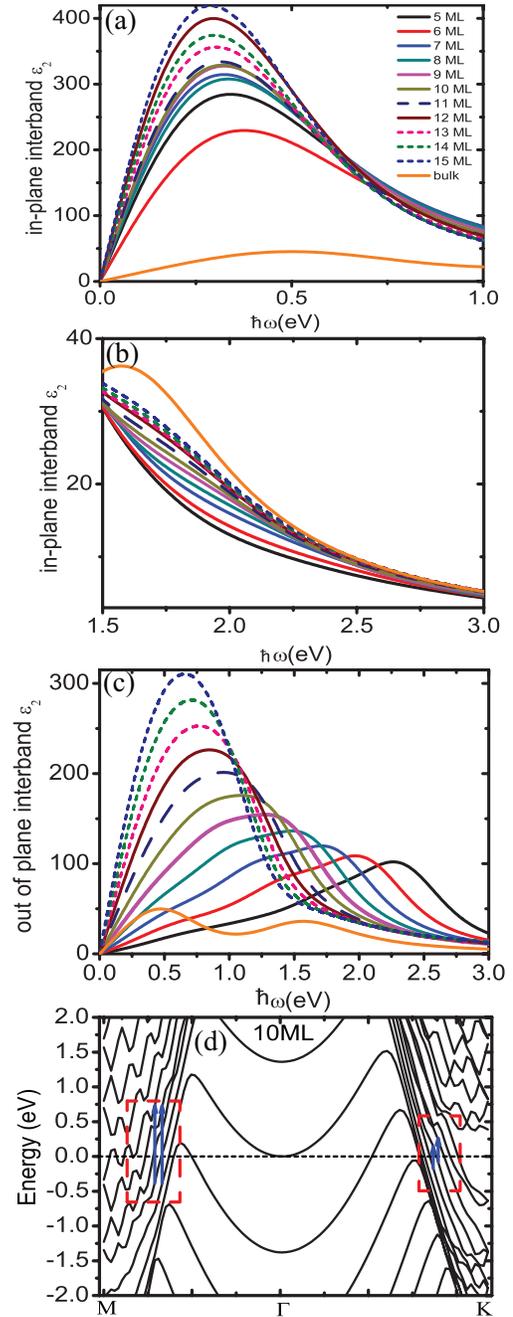}
\caption{\label{Fig4} Imaginary part of interband in-plane
dielectric function of Al(111) films with the thickness from 5~ML to
15~ML (a) at low energy range  and (b) relatively high energy range.
(c)The imaginary part of interband out-of-plane dielectric function
. For comparison, the bulk curve of dielectric function is also
plotted in (a)-(c). (d)The band structure of 10~ML AL(111) film in
order to track the optical transitions responsible for the peaks in
its dielectric functions. The short arrows indicate the possible
transitions between two parallel bands of $\sim$0.3~eV energy
difference and the long arrows indicate the possible transitions
between two parallel bands of $\sim$1.0~eV energy difference.}
\end{figure}

Next, we investigated QSE on the interband contribution to the
dielectric function. In Fig. 3 we plotted the imaginary part of the
in-plane $\varepsilon_2$ versus film thickness for three different
photon energies ($\hbar \omega=$0.5~eV, 1.5~eV and 3.0~eV). The
dashed line in each plot represents the corresponding bulk value at
the same energy. We can see that the value at $\hbar \omega=$0.5~eV
is much larger than that in the bulk \cite{Alplas1}. However, at
$\hbar \omega=$1.5~eV and $\hbar \omega=$3.0~eV the value at large
thickness is almost the same as the bulk value. This is due to the
additional interband transitions in the film from the quantum-well
states but absent in the bulk. It is inversely proportional to
$\omega^2$ from equation (\ref{inter2}), giving rise to larger value
in the low energy regime (this trend will be more clear from the
dielectric function plot in Fig. 4 below). It's interesting to also
note that at $\hbar \omega=$1.5~eV and $\hbar \omega=$3.0~eV the
dielectric function is smaller for film thickness less than 15~MLs
than that of bulk (see Fig. 3(b-c)). This lower dielectric constant
will proportionally result in lower absorption loss. It suggests to
us that ultrathin film can be used to achieve lower loss plasmonic
devices in visible and UV frequency. In the film with thickness less
than 4~MLs, the dielectric constant is significantly lower than the
bulk value, indicating much stronger QSE modulation due to the
strong quantum confinement perpendicular to the surface, where
absorption loss will be the least. However, for such ultrathin film,
the slab plasma frequency is larger as shown in Fig. 2(a) which will
leads to larger damping induced energy loss. In Fig. 3(d-e) the
out-of-plane interband $\varepsilon_2$ are showed. It is also much
larger than the bulk at low photon energy ($\hbar \omega=$0.5~eV)
but very close to the bulk at higher energies ($\hbar \omega=$1.5~eV
and $\hbar \omega=$3.0~eV). Therefore the symmetry breaking in
ultrathin film will have significant influence on both the in-plane
and out-of-plane $\varepsilon_2$.

Finally in Fig. 4 we plotted the $\varepsilon_2$ for film thickness
from 5~MLs to 15~MLs together with bulk interband $\varepsilon_2$
for comparison. Fig. 4(a) and Fig. 4(b) show the low energy part
($\leq 1.0~eV$) and high energy part (($\geq$ 1.0~eV)) of the
in-plane interband $\varepsilon_2$, respectively. In low energy
range it is up to 10 times larger than the bulk value. This is
consistent with the decreased slab plasma frequency in Fig. 2(a)
which will result in larger interband $\varepsilon_2$ in some
frequency range because of the \emph{f}-sum rule in equation
(\ref{fsum}). Also the film in-plane dielectric function has a
red-shifted peak position from 0.5~eV in bulk Al. For example, the
peak is moved to $\sim$0.28~eV for thickness of 15~MLs. Similarly in
Fig.4(c) the out-of-plane interband dielectric function
$\varepsilon_2$ exhibits redshifted and enhanced peak compared to
the peak of the bulk at 1.5~eV. The origin of this redshifted and
enhanced peak is related to the band structure of the film. We have
plotted the band structure of 10~ML film as a representative
thickness in Fig. 4(d). Around Fermi energy parallel bands with
energy difference of ~0.3 eV and ~1.0~eV are found, which are
indicated by short and long arrows, respectively. The peak around
0.3~eV in the in-plane dielectric function plot for 10~ML film in
Fig. 4(a) will be from the transitions in parallel bands with energy
difference of ~0.3~eV. The peak around 1.0~eV in the out-of-plane
dielectric function plot for 10~ML film in Fig. 4(c) will be from
the transitions in parallel bands with energy difference of ~1.0~eV.
The higher in-plane interband dielectric function $\varepsilon_2$
than the bulk couterpart indicates the larger absorption loss in the
low energy range. On the other hand, from Fig. 4(b) we found in
between 1.5~eV and 2.5~eV film interband $\varepsilon_2$ instead
becomes larger than the bulk $\varepsilon_2$. Therefore within this
regime, lower absorption loss in plasmonics device is expected. For
photon energy greater than 2.5~eV, the difference between films of
different thickness and bulk is insignificantly small.

\section{Conclusion} We performed DFT calculations of the dielectric
function of Al(111) ultrathin films of thickness from 1 to 39~MLs
with RPA. Both the intraband and inerband contributions to the total
dielectric function were shown to be modulated by QSE through the
calculation of slab plasma frequency and the imaginary part of the
dielectric function, which are proportional to incident photon
energy loss with free carrier damping mechanism and electron optical
transition induced energy absorption mechanism, respectively. The
key findings are: (1) in the low photon energy ($\leq$ 1~eV) range,
the interband transition energy loss is significantly enhanced
compared to bulk value; however, (2) in a narrow relatively high
energy range (1.5~eV to 2.5~eV) the in-plane interband transition
energy is reduced and overall the thinner film is, the less the
interband absorption of in-plane polarized light is for films less
then 15~MLs. Our result may shed light on the search for low
absorption loss plasmonics material through electronically tuning
the band structure of metallic film by using different thickness.

This work was supported by NSF MRSEC (Grant No. DMR-1121252) and
DOE-BES (Grant No. DE-FG02-04ER46148). We thank the CHPC at
University of Utah and NERSC for providing the computing resources.


\end{document}